\documentclass[runningheads]{llncs}
\usepackage{graphicx}
\usepackage{amsmath,amssymb} 
\usepackage{color}

\usepackage{adjustbox}

\usepackage{subfigure}

\usepackage{mathtools}
\usepackage{comment}

\begin{document}
\title{Photorealistic Facial Synthesis in the Dimensional Affect Space} 

\titlerunning{Photorealistic Facial Synthesis in the Dimensional Affect Space}

\author{Dimitrios Kollias\inst{1} \and
Shiyang Cheng\inst{1}
\and Maja Pantic\inst{1} \and
Stefanos Zafeiriou\inst{1,2}}
%
\authorrunning{D. Kollias, S. Cheng and S. Zafeiriou}
%

\institute{ Department of Computing, Imperial College London, UK\and
 Centre for Machine Vision and Signal Analysis, University of Oulu, Finland
\email{\{dimitrios.kollias15,shiyang.cheng11,s.zafeiriou\}@imperial.ac.uk}}
\maketitle              
\begin{abstract}
This paper presents a novel approach for synthesizing facial affect, which is based on our annotating 600,000 frames of the 4DFAB database in terms of valence and arousal. 
The input of this approach is a pair of these emotional state descriptors and a neutral 2D image of a person to whom the corresponding affect will be synthesized.
Given this target pair, a set of 3D facial meshes is selected, which is used to build a blendshape model and generate the new facial affect. 
To synthesize the affect on the 2D neutral image, 3DMM fitting is performed and the reconstructed face is deformed to generate the target facial expressions. Last, the new face is rendered into the original image.
Both qualitative and quantitative experimental studies illustrate the generation of realistic images, when the neutral image is sampled from a variety of well known databases, such as the Aff-Wild, AFEW, Multi-PIE, AFEW-VA, BU-3DFE, Bosphorus.

\end{abstract}

\section{Introduction}

Rendering photorealistic facial expressions from single
static faces while preserving the identity information is an open research topic which has 
significant impact on the area of affective computing.
Generating faces of a specific person with different facial
expressions can be used to various applications including face
recognition~\cite{cao2018vggface2}~\cite{parkhi2015deep}, face verification~\cite{sun2014deep}~\cite{taigman2014deepface}, emotion prediction~\cite{kollias2016line}~\cite{kollias2017adaptation}~\cite{1809.04359}, expression database generation, augmentation and entertainment. 

This paper describes a novel approach that takes an arbitrary face
image with a neutral facial
expression and synthesizes a new face image of the same
person, but with a different expression, generated according to a dimensional emotion representation model. 
This problem cannot be tackled using small databases
with labeled facial expressions, because it would be really difficult to disentangle facial expression and identity information through them.
Our approach is based on the analysis of a large 4D facial database, the 4DFAB \cite{cheng4dfab}, which we appropriately annotated and used for facial expression synthesis on a given subject's face. 
A dimensional emotion model, in terms of the continuous variables valence (i.e., how positive or negative is an emotion) and arousal (i.e., power of the activation of the emotion) \cite{whissell1989dictionary} \cite{russell1978evidence}, has been used to annotate the large amounts of facial images, since this model can represent, not only primary, extreme expressions, but also subtle expressions which are met in everyday human to human, or human to machine interactions.

Section 2 refers to related work that has been published with reference to facial expression synthesis. Section 3 presents the proposed approach for generating facial affect. We describe the annotation and use of the 4DFAB database, and provide the pipeline of our approach in detail. In Section 4, we provide an evaluation of the Valence - Arousal discretization and modeling procedure. Then, we synthesize facial affect on a variety of neutral faces from ten different databases (annotated either using a categorical or dimensional emotion model). By using augmented data of faces from two in-the-wild databases, we train a deep neural network to predict the valence and arousal values in these databases. Experimental results show that the proposed approach manages to synthesize photorealistic facial affect, which can be used to improve the accuracy of valence and arousal prediction. Conclusions and future work are presented in Section 5.

\section{Related Work}
In the past several years, facial expression synthesis has been an active research topic. All facial expression synthesis methods that were proposed in the past two decades were roughly split into two categories.
The first category is mainly using computer graphics
techniques in order to directly warp input faces to target expressions \cite{zhang2006geometry} \cite{yang2012facial} \cite{yeh2016semantic} or re-use sample patches of existing images \cite{mohammed2009visio}. The second one synthesizes images with attributes that are predefined \cite{ding2017exprgan}  \cite{susskind2008generating} through the creation of generative models.
For the first category, a lot of research efforts have been
devoted to finding the correspondence between the target images and existing facial
textures. Earlier approaches mostly generated
new expressions by either compositing face
patches from an existing expression database \cite{mohammed2009visio} \cite{jonze1999being}, or warping face images via optical flow \cite{yang2012facial} \cite{yang2011expression} and feature correspondence \cite{theobald2009mapping}, or creating fully textured 3D facial
models \cite{pighin2006synthesizing} \cite{blanz2003reanimating}.
In particular, \cite{yeh2016semantic} proposed to learn the optical flow using a
variational autoencoder. Although this kind of methods can
usually produce realistic images with high resolution, the
elaborated complex processes often result in highly expensive
computations.
These works have shown either how to
synthesize facial expressions on virtual agents \cite{zhang2010facial},
or
how to transfer facial
expressions between different subjects, i.e., facial reenactment \cite{thies2015real}. However, synthesizing accurately a wide variety of facial expressions on arbitrary real faces is considered an open problem and has much room for improvement.

Due to this difficulty, the second category of methods has initially focused on using
deconvolutional neural networks (DeCNNs) \footnote{https://zo7.github.io/blog/2016/09/25/generating-faces.html} or deep belief nets (DBNs) \cite{susskind2008generating}, 
generating faces through interpolation of the facial images in their training
set. This, however, makes them inherently unsuited for facial expression
generation in the case of unseen subjects. With the recent development of Generative Adversarial
Networks (GANs) \cite{goodfellow2014generative}, image editing has migrated from pixel-level
manipulations to semantic-level ones. GANs
have been successfully applied to face image editing, for modification of
facial attributes \cite{yan2016attribute2image} \cite{ghodrati2015towards}, age
modeling \cite{zhang2017age} and pose adjustment \cite{huang2017beyond}. These methods generally use the encoder of the GAN to find a low-dimensional representation of the face image in a latent space, manipulate the latent vector and then decode it to generate the new image. 

Popular approaches
shift the latent
vector along a direction corresponding to semantic
attributes \cite{larsen2015autoencoding} \cite{yeh2016semantic}, or 
concatenate attribute labels with it \cite{zhang2017age} \cite{yan2016attribute2image}.
Adversarial discriminator networks are used, either at the
encoder to regularize the latent space \cite{makhzani2015adversarial}, or at the decoder to
generate blur-free and realistic images \cite{larsen2015autoencoding} or at both encoder and decoder, such as the Conditional Adversarial
Autoencoder. All of these approaches require
large training databases so that identity information can be
properly disambiguated. Otherwise, when presented with an
unseen face, the network tends to generate faces which look
like the “closest” subject in the training database. It has been proposed to handle this problem by warping
images, rather than generating them from the latent vector \cite{yeh2016semantic}. This approach achieves a high interpolation quality, but requires that the input expression is known and fails when generating facial expressions that are “far apart,” e.g. angry faces from smiling ones.
Moreover, it is hard to take fine-grain control of the synthesized images, e.g., widen the smile or narrow the eyes.

The proposed approach has quite a few novelties. First of all, it is the first time, to the best of our knowledge, that the dimensional model of affect is taken into account when synthesizing images. All other models are producing synthesized images according to the seven basic, or a few more, expressions. Our approach, as verified in the experimental section of this paper, produces a large number of different expressions given a valence and arousal pair of values in the continuous 2D domain. Also, it is the first time that a 4D face database is annotated in terms of valence and arousal and is then used for affect synthesis.
What is more, until now, there has not been any attempt to use the blendshape models like we propose for the synthesis of the data. 
Finally, the proposed approach works well, when presented with a neutral image  either from a controlled or from an in-the-wild database and with different head poses of the person appearing in that image.     

\section{The Proposed Approach}
\subsection{The 4DFAB database}
\label{subsec:4dfab}
The 4DFAB database~\cite{cheng4dfab} is the first large scale 4D face database designed for biometrics applications and facial expression analysis. It consists of 180 subjects (60 females, 120 males) aging from 5 to 75. 4DFAB was collected over a period of 5 years under four different sessions, with over 1,800,000 3D faces. The database was designed to capture articulated facial actions and spontaneous facial behaviors, where spontaneous expressions are elicited by emotional video clips watching. In this paper, we use all the 1,580 spontaneous expression sequences for our emotion analysis and synthesis; these sequences cover a wide range of expressions as defined in~\cite{compound_expr_pnas}.

\begin{figure*}[h!]
\centering
\adjincludegraphics[height=5cm,width=7cm,trim={{.12\totalheight} {.12\totalheight} {.12\totalheight} {.12\totalheight}}]{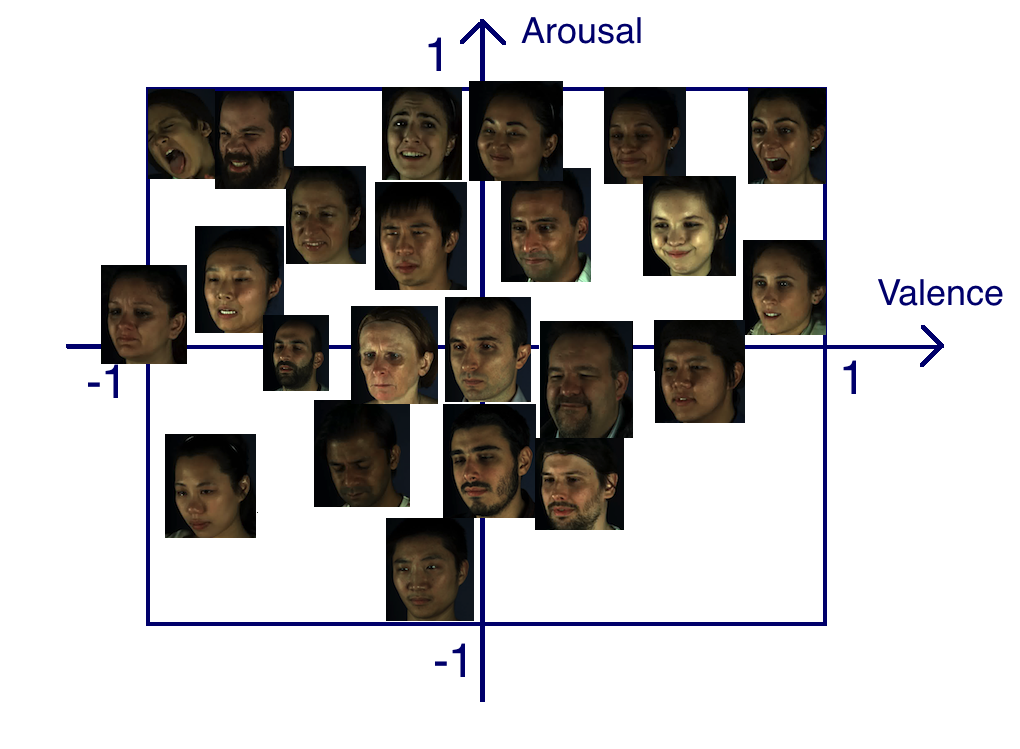}  
\caption{The 2D Valence-Arousal Space and some representatives frames of 4DFAB}
\label{2d_wheel}
\end{figure*}

To be able to develop the novel expression synthesis method, we annotate these dynamic 3D sequences (over 600,000 frames), in terms of valence and arousal emotion dimensions, using the tool described in \cite{zafeiriou-affw}. 
Valence and arousal values range in [-1,1]. Examples are shown in Fig.~\ref{2d_wheel}. In the rest of the paper, when we refer to the 4DFAB database, we mean the 600,000 frames which are annotated with categorical expressions, as well as 2D valence and arousal (V-A) emotion values. 

As each 3D face in 4DFAB differs in the number, as well as topology of vertex, we need to first correlate all these meshes to an universal coordinate frame - namely a 3D face template. This step is usually called establishing dense correspondence. We follow the same UV-based registration approach in~\cite{cheng4dfab} to bring all the 600,000 meshes into full correspondence with the mean face of LSFM~\cite{booth2018large}. As a result, we create a new set of 600,000 3D faces that share identical mesh topology, while maintaining their original facial expressions; we will use them as our 3D facial expression gallery for the facial affect synthesis.

\subsection{The methodology pipeline} 
\begin{figure*}
\centering
\adjincludegraphics[width=0.99\linewidth]{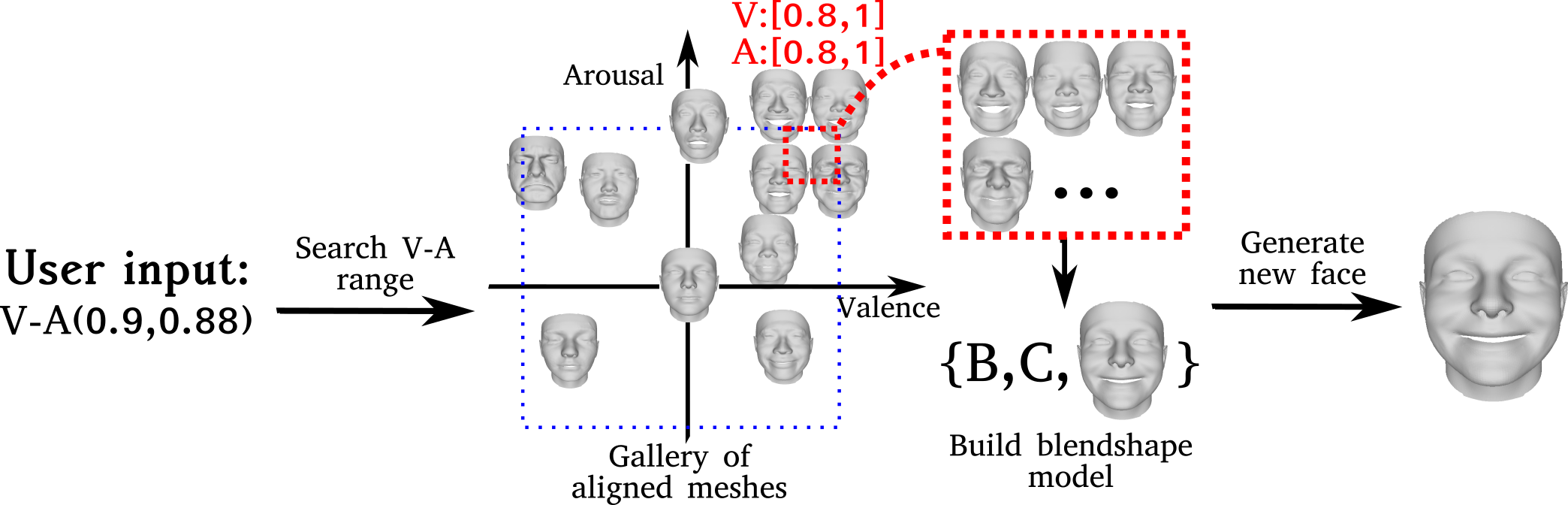}\\
(a) Generate new facial affect, given a target V-A pair.\\
\adjincludegraphics[width=0.9\linewidth]{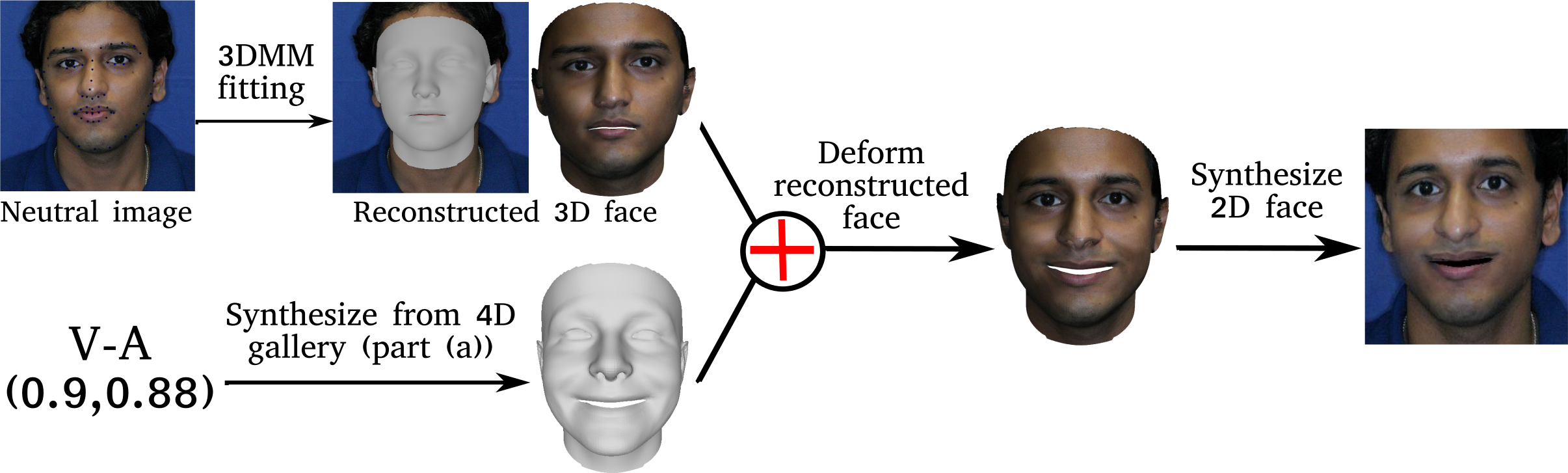}\\
(b) synthesize facial affect on a 2D neutral face.\\
\caption{Two main parts in our facial affect synthesis framework: (a) generating new facial affect from our 4D face gallery, given a target V-A value pair provided by the user; (b) synthesizing the facial affect (from part (a)) on an arbitrary 2D neutral face.}
\label{framework}
\end{figure*}

The main novelty and contribution of this paper comes from the development of a fully automatic facial affect synthesis framework (depicted in Fig. \ref{framework}). In the first part (Fig.~\ref{framework}(a)), assuming that the user inputs a target V-A pair, we aim at generating semantically correct 3D facial affect from our 4D gallery. There are two key stages in this pipeline. The first includes the data selection from the 4D face gallery and the utilization of these data. To this end, we discretize the 2D Valence-Arousal (V-A) Space into 100 classes (see Fig.~\ref{blend_mean} for visualization). Each class contains aligned meshes that are associated with the corresponding V-A pairs; all these V-A pairs lie within the area of this class. Therefore, when a user provides us with a V-A pair, we find its class and retrieve the data belonging to this class. We then build a blendshape model using these data and compute the mean face. Eventually, using this blendshape model, we can generate an unseen 3D face with affect. The details of this part are described in Fig. \ref{framework}(a) and Section~\ref{subsec:generate_3D_affect}.

Fig. \ref{framework}(b) describes the procedure of synthesizing a new facial affect to an arbitrary 2D face. As described previously, given a target V-A pair, we create an unseen expressive face without any identity, gender and age information. In this part, we want to transfer the affect of this expressive face to the face of another person, after which, we render a 2D expressive face without loss of identity. Three processing steps are needed to achieve this goal. The first is to perform 3DMM fitting~\cite{booth2017itw3dmm} to estimate the 3D shape of target face. The second step is to transfer the facial affect from synthetic 3D face to the reconstructed 3D face. Finally, we rasterize the new 3D face with affect to the original image frame. We will describe this procedure in details in Section~\ref{subsec:syn_to_2d}.

\subsection{Generation of new 3D facial affect from 4DFAB}
\label{subsec:generate_3D_affect}
\subsubsection{Discretizing the 2D Valence-Arousal Space}
\begin{figure*}
\centering
\adjincludegraphics[height=6cm,width=12cm]{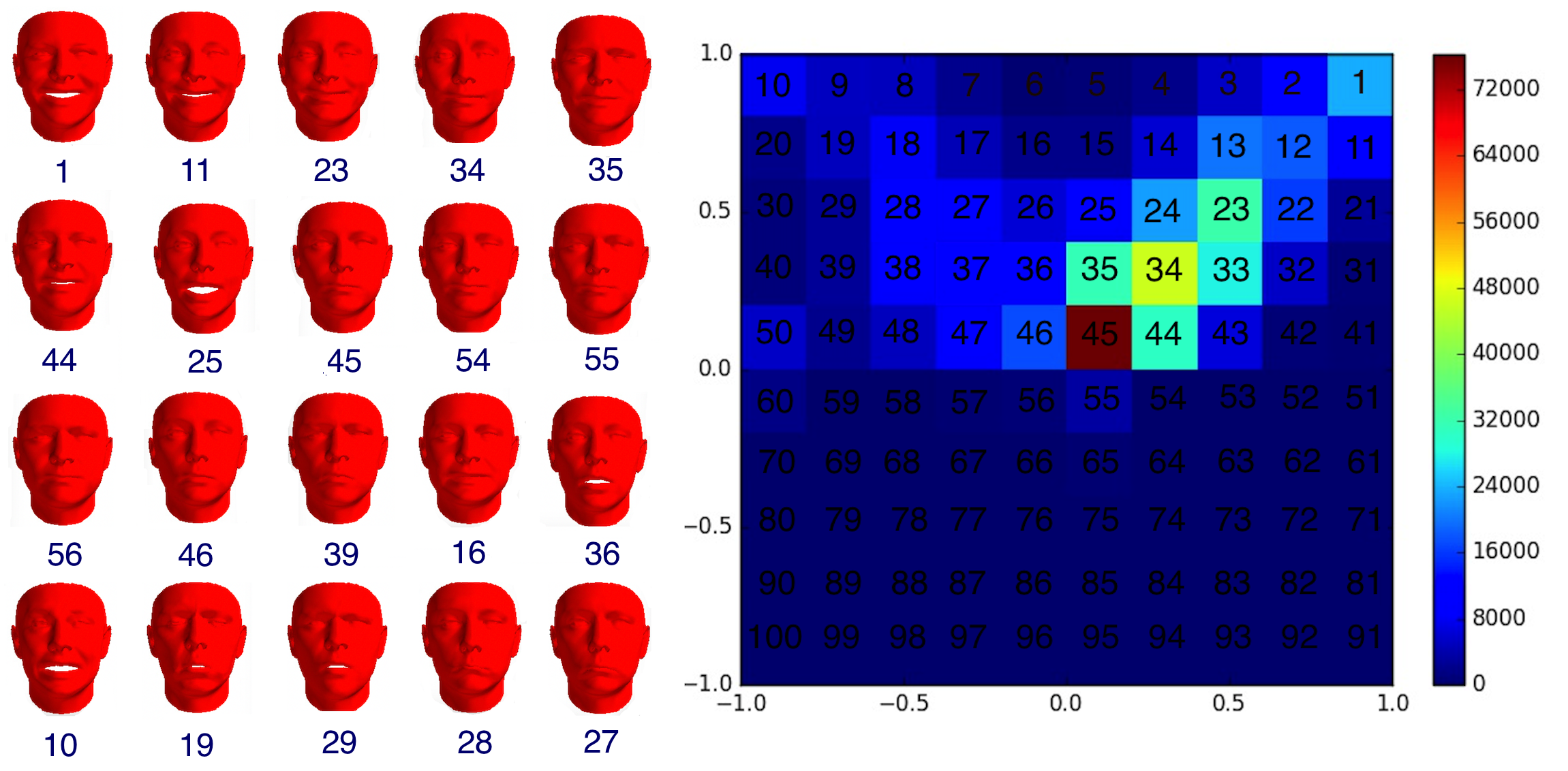} 
\caption{The mean shapes of our blendshape models and their corresponding areas in the 2D Valence-Arousal Space, which is shown as a 2D histogram of annotations of the 4DFAB database.}
\label{blend_mean}
\end{figure*}

At first, we discretize the 2D Valence-Arousal Space into 100 classes, with each one covering a square of size $0.2 \times 0.2$ and including a sufficient number of data. Although the number of classes can be increased to further categorize the facial affect, it might not provide a better result. This is because, if each class contained few examples, it would be more likely that the identity information is incorporated. However, our synthetic facial affects should only describe the expression associated with the designated V-A value pair, rather than any of the identity, gender and age information. Fig. \ref{blend_mean} shows on the right side the histogram of annotations (of 4DFAB database) of the discretized Valence-Arousal Space and on the left side the corresponding mean blendshapes of various classes of this Space. 
Expression blendshape models provide an effective way to parameterize facial behaviors and are frequently used in many computer vision applications. We choose to build the localized blendshape model~\cite{neumann2013sparse} to describe our selection of V-A examples. For each 3D mesh, we subtracted it from the neutral mesh of the corresponding sequence and created a set of $m$ difference vectors $\mathbf{d}_i \in \mathbb{R}^{3n}$ which were then stacked into a matrix $\mathbf{D}=[\mathbf{d}_1, ..., \mathbf{d}_m] \in \mathbb{R}^{3n \times m}$, where $n$ is number of vertices in our mesh. Afterwards, a variant of sparse Principal Component Analysis (PCA) was applied to our data matrix $\mathbf{D}$ to identify sparse deformation components $\mathbf{C} \in \mathbb{R}^{h \times 1}$:
\begin{equation}
    \arg \min \left \| \mathbf{D} - \mathbf{B} \mathbf{C} \right \|_F^2 + \Omega \left ( \mathbf{C} \right ) \; \; \; \textup{s.t.} \; \mathcal{V}\left (  \mathbf{B} \right ), 
\label{eq:blendshape_spca}
\end{equation}
here, the constraint $\mathcal{V}$ can be either $\max \left ( \left | \mathbf{B}_{k} \right | \right ) = 1, \; \forall k$ or $\max \left ( \mathbf{B}_{k} \right ) = 1, \; \mathbf{B} \geq 1, \; \forall k$, where $\mathbf{B}_k \in \mathbb{R}^{3n \times 1}$ denotes the $k^{th}$ components of sparse weight matrix $\mathbf{B} = [ \mathbf{B}_1, \cdots, \mathbf{B}_h]$. The selection of these two constraints depends on our actual usage; the major difference is that the latter one allows negative weights and therefore enables deformation towards both directions, which is useful for describing shapes like muscle bulges. The regularization of sparse components $\mathbf{C}$ is performed with $\ell1 / \ell2$ norm~\cite{Wright2009,Bach2012}. To permit more local deformations from the model, additional regularization parameters were added into $\Omega \left ( \mathbf{C} \right )$. To solve for the optimal $\mathbf{C}$ and $\mathbf{B}$, an iterative alternating optimization is employed, please refer to~\cite{neumann2013sparse} for more details.

\subsection{Facial affect synthesis for arbitrary 2D image}
\label{subsec:syn_to_2d}
Given a facial expression synthesis based on the valence-arousal value pair, we aim at modifying the face in an arbitrary 2D image and generating a new facial image with affect. This procedure consists of three steps: (1) fit a 3D morphable model on the image; (2) generate facial affect on the reconstructed 3D face; (3) blend the new face into the original image. Specifically, we started by performing a 3DMM fitting~\cite{booth2017itw3dmm} on a 2D facial image, and retrieved a reconstructed 3D face with the texture sampled from the original image. Next, we calculated the facial deformation by subtracting the synthetic face with the LSFM template, and imposed this deformation on the reconstructed mesh. This far, we have generated a new 3D face with certain affect; the last step would be rendering it back to the original 2D image, where a Poisson image blending~\cite{Perez2003Siggraph} is employed to produce a natural and realistic result.

\section{Experimental Study}

\subsection{Discovering shared information between 3D data and Valence-Arousal}
\label{subsec:shared_info}
In the first experiment, we wanted to prove the validity of our Valence-Arousal modeling and synthesis approach. This could be verified by showing that there is shared information between our 3D data and Valence-Arousal through a correlation analysis.

Due to the high volume and dimensionality of our 3D data, it is intractable to directly perform typical correlation analysis. Hence, we first built a powerful expression blendshape model using the apex frames of posed expression sequences from the 4DFAB; in total, 12,000 expressive 3D meshes are selected, with 2,000 for each of the six basic expressions. Then, we projected our 3D data to its subspace and retrieved the sparse representations for future analysis. We experimented with different number of components (i.e. 84, 150, 200, 300, 500) of the blendshape model to select the best configuration. 

Next, we split the data into 2 sets: the training and the test set, containing 480,000 and 120,000 frames respectively, in a subject independent manner, meaning that one person could only appear in the training or test set, but not on both of them. As we have found a compact representation of our data, Canonical Correlation Analysis (CCA)~\cite{eps259225} can be performed on the training set and their corresponding valence and arousal values. CCA is a shared-space component analysis method, which recovers the loadings to project two data matrices on a subspace where the linear correlation is maximized. This can be interpreted as discovering the shared information conveyed by all the data (or views).

After CCA, we reduced the dimensions of our data to 2. Then, on the training set, we performed Support Vector Regression (SVR) \cite{basak2007support} with Radial Basis Function (RBF) kernel to map those 2 dimensions to the valence and arousal values. In order to examine whether our 3D data highly correlate to the Valence-Arousal labels, we predicted the V-A values of the test data using the aforementioned models (CCA and SVR), and compare the predictions with our annotated V-A labels. This comparison was performed with respect to two criteria: Concordance Correlation Coefficient and the usual Mean Squared Error. The Concordance Correlation Coefficient (CCC) can be defined as follows:
\begin{equation} \label{eq:6}
\rho_c = \frac{2 s_{xy}}{s_x^2 + s_y^2 + (\bar{x} - \bar{y})^2},
\end{equation}
where $s_x$ and $s_y$ are the variances of the ground truth and predicted values of the regression respectively, $\bar{x}$ and $\bar{y}$ are the corresponding mean values and $s_{xy}$ is the respective covariance value.

Table \ref{cca} shows those two criteria for the test set when we keep different numbers of principal components for our expression blendshape model. We can observe that with 200 components, highest correlation between the data and V-A labels was achieved, as well as lowest prediction error. By selecting this value, we ensured that the proposed synthesis approach is valid.

\setlength{\tabcolsep}{4pt}
\begin{table}
\begin{center}
\caption{CCC and MSE evaluation of valence \& arousal predictions on the test set when we keep different number of principal components in PCA }
\label{cca}
\begin{tabular}{ |c||c|c|c|c|  }
\hline 
\multicolumn{1}{|c||}{\begin{tabular}{@{}c@{}} No. of principal \\ components to keep \end{tabular}} & \multicolumn{2}{|c|}{CCC} & \multicolumn{2}{|c|}{MSE}  \\ 
\hline
     & Valence & Arousal & Valence & Arousal  \\
\hline
84  & 0.63 & 0.68 & 0.107 &0.046\\
\hline
150 & 0.65 & 0.68 & 0.099 &0.041\\
\hline
\textbf{200} & \textbf{0.66} & \textbf{0.69} & \textbf{0.097} &\textbf{0.040}\\
\hline
300 & 0.35 & 0.30 & 0.127 &0.058\\
\hline
500 & 0.31 & 0.22 & 0.129 &0.061\\
\hline
\end{tabular}
\end{center}
\end{table}
\setlength{\tabcolsep}{1.4pt}
\subsection{Databases used for affect synthesis evaluation}

To evaluate our facial affect synthesis method in different scenarios (e.g. controlled laboratory environment, uncontrolled in-the-wild setting), we utilized neutral facial images from as many as 10 databases. 

\medskip
\noindent \textbf{1) Multi-PIE}~\cite{gross2010multi}: It contains 755,370 images (3072x2048) of 337 people. Pose, illumination, and expression are the key factors of the database. 15 view points, 19 illuminations and 7 expressions are recorded in a controlled environment. 

\noindent \textbf{2) Face place:} 
This database \footnote{Stimulus images courtesy of Michael J. Tarr, Center for the Neural Basis of Cognition and Department of Psychology, Carnegie Mellon University, http://www.tarrlab.org/} contains photographs of many different individuals in various types of disguises, such that, for each individual, there are multiple photographs in which hairstyle and/or eyeglasses have been changed/added. It consists of 1,284 images of Asian, 937 images of African-American, 3,362 images of Caucasian, 494 images of Hispanic and 497 images of multiracial people. All images show posed expression.

\noindent \textbf{3) 2D Face Sets:} 
We used 3 subsets from the 2D Face Sets database\footnote{http://pics.stir.ac.uk}.

\noindent\underline{Iranian women}: It consists of 369 color images (1200x900) of 34 women. People display mostly smile and neutral expression in each of five poses.

\noindent\underline{Nottingham scans}: It has 100 monochrome images (50 men, 50 women) in neutral and frontal pose. The image resolution varies from 358x463 to 468x536. 

\noindent\underline{Pain expressions}: It consists of 599 color images (720x576) of 13 women and 10 men. They usually display two of the six basic emotions (anger, disgust, fear, sad, happy, surprise) plus pain 10 expressions. Profile neutral and 45 degrees images are available.

\noindent \textbf{4) FEI:} 
The FEI database \cite{thomaz2010new} is a Brazilian face database that contains a set of face images taken between June 2005 and March 2006. 200 individuals were recorded, and each one has 14 images, resulting in 2,800 images of size 640x480. All images were color and taken against a white background in an upright frontal position with profile rotation of up to 180$^\circ$. The subjects are mostly students and staff at FEI, between 19 and 40 years old with distinct appearance, hairstyle and adorns. The number of male and female subjects are both 100.

\noindent \textbf{5) Aff-Wild:} 
Aff-Wild~\cite{1804.10938}~\cite{zafeiriou-affw} consists of 298 Youtube videos, with $1,200,000$ frames in total. The length of each video varies from 10 seconds to 15 minutes. These videos contain spontaneous facial behaviors elicited by a variety of stimuli in arbitrary recording conditions. There are 200 subjects (130 males and 70 females) from different ethnicities. Aff-Wild serves as the benchmark of the first Affect-in-the-wild Challenge\footnote{https://ibug.doc.ic.ac.uk/resources/first-affect-wild-challenge}~\cite{kollias2017recognition}. For each video, there are 8 annotators to annotate the valence and arousal, in the range of [$-1 $, $+1 $].

\noindent \textbf{6) AFEW 5.0:} 
This database is a dynamic facial expressions corpus (used in EmotiW Challenge 2017~\cite{dhall2017individual}) consisting of 1,809 nearly real world scenes from movies and reality TV shows. There are over 330 subjects aging from 1 to 77. The database is split into three sets: training (773 videos), validation (383 videos) and test set (653 videos). It is a challenging database because both training and validation sets are mainly from the movies, while 114 out of 653 test videos are from TV. Annotations of neutral and 6 basic expressions are provided. 

\noindent \textbf{7) AFEW-VA:} 
Recently, a part of the AFEW database has been annotated in terms of Valence and Arousal, thus creating the AFEW-VA~\cite{kossaifi2017afew} database. It includes 600 video clips selected from films with real-world conditions, i.e., occlusions, illumination and body movements. The length of each video ranges from around 10 frames to over 120 frames. This database consists of per-frame annotations of V-A. In total, more than 30,000 frames were annotated for affect prediction of V-A, using discrete values in the range of [$-10 $, $+10 $]. 

\noindent \textbf{8) BU-3DFE:} 
BU-3DFE database~\cite{yin20063D} is the first 3D facial expression database, which includes 2,500 expressive meshes from 100 subjects (56 females, 44 males) with age from 18 to 70. The subjects are from various ethnic/racial ancestries. They recorded 6 articulated expressions (happiness, disgust, fear, angry, surprise and sadness) with 4 intensities; also, there is a neutral 3D scan per subject.

\noindent \textbf{9) Kinect Fusion ITW:} 
The KF-ITW database~\cite{booth2017itw3dmm} is the first Kinect 3D database captured under relatively unconstrained conditions. This database consists of 17 different subjects performing some expressions (neutral, happy, surprise) under various illumination conditions.

\noindent \textbf{10) Bosphorus:} 
The Bosphorus database \cite{savran2008bosphorus} consists of 105 subjects in various poses, expressions and occlusion conditions. 18 men had beard/moustache and 15 others had short facial hair. There are 60 men and 45 women, they are mostly between 25 and 35. Majority of them are Caucasian. 27 professional actors/actresses are incorporated in the database. The number of total face scans is 4,652, each scan has been manually labeled with 24 facial landmarks.

\subsection{Qualitative evaluation of the facial affect synthesis}

We used all the above-mentioned databases to supply the proposed approach with 'input' neutral faces. We synthesized the emotional state of specific V-A value pairs for these images. One important task during this facial affect synthesis procedure is to preserve identity, age and gender of the original face. Instead of finding the closest matching sample (or K-nearest samples) for the given V-A pair, we categorized our 3D data based on the 2D Valence-Arousal Space (as shown in Fig. \ref{blend_mean}) and employed the mean expression of the area that contains the target V-A pair.

\begin{figure*}
\centering
\adjincludegraphics[width=1\linewidth]{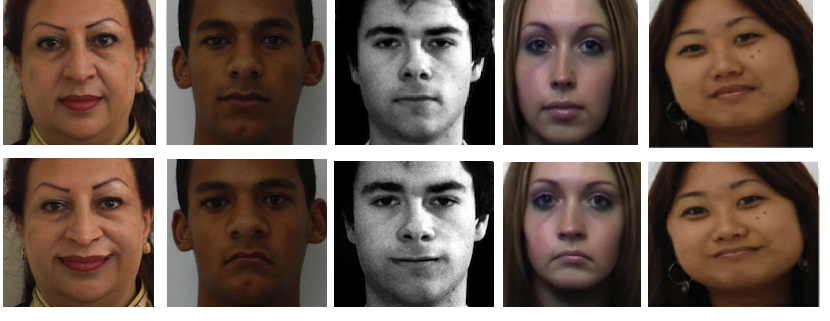}\\
(a) \\
\adjincludegraphics[width=1\linewidth]{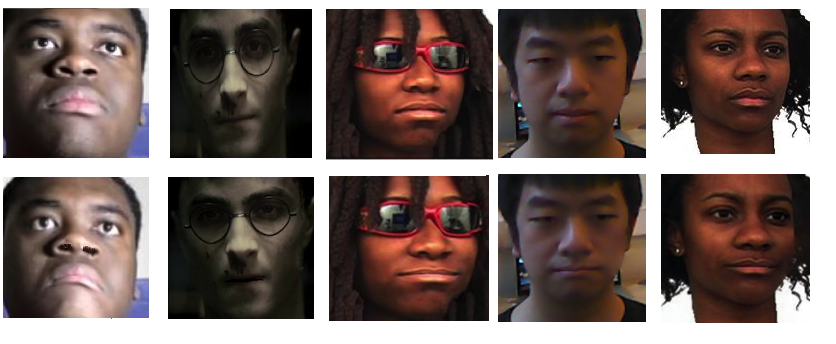}\\
(b) \\
\adjincludegraphics[width=1\linewidth]{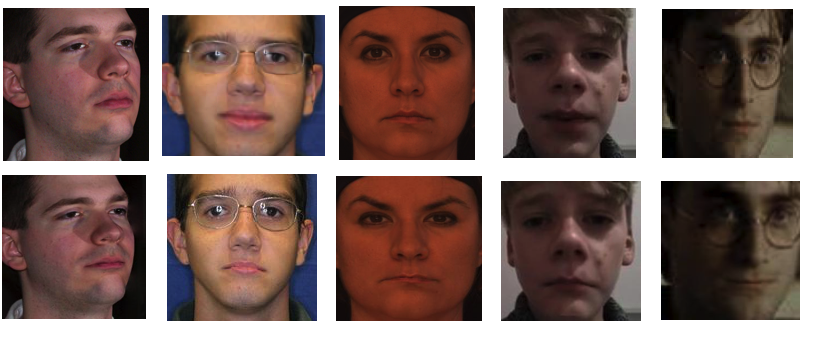}\\
(c) \\
\caption{\textbf{(a)-(c).} Synthesis of facial affect across all databases: on top rows are the neutral and on the bottom are the corresponding synthesized images.}
\label{results_all_dbs}
\end{figure*}

\begin{figure*}
\centering
\adjincludegraphics[height=15cm,width=12.2cm]{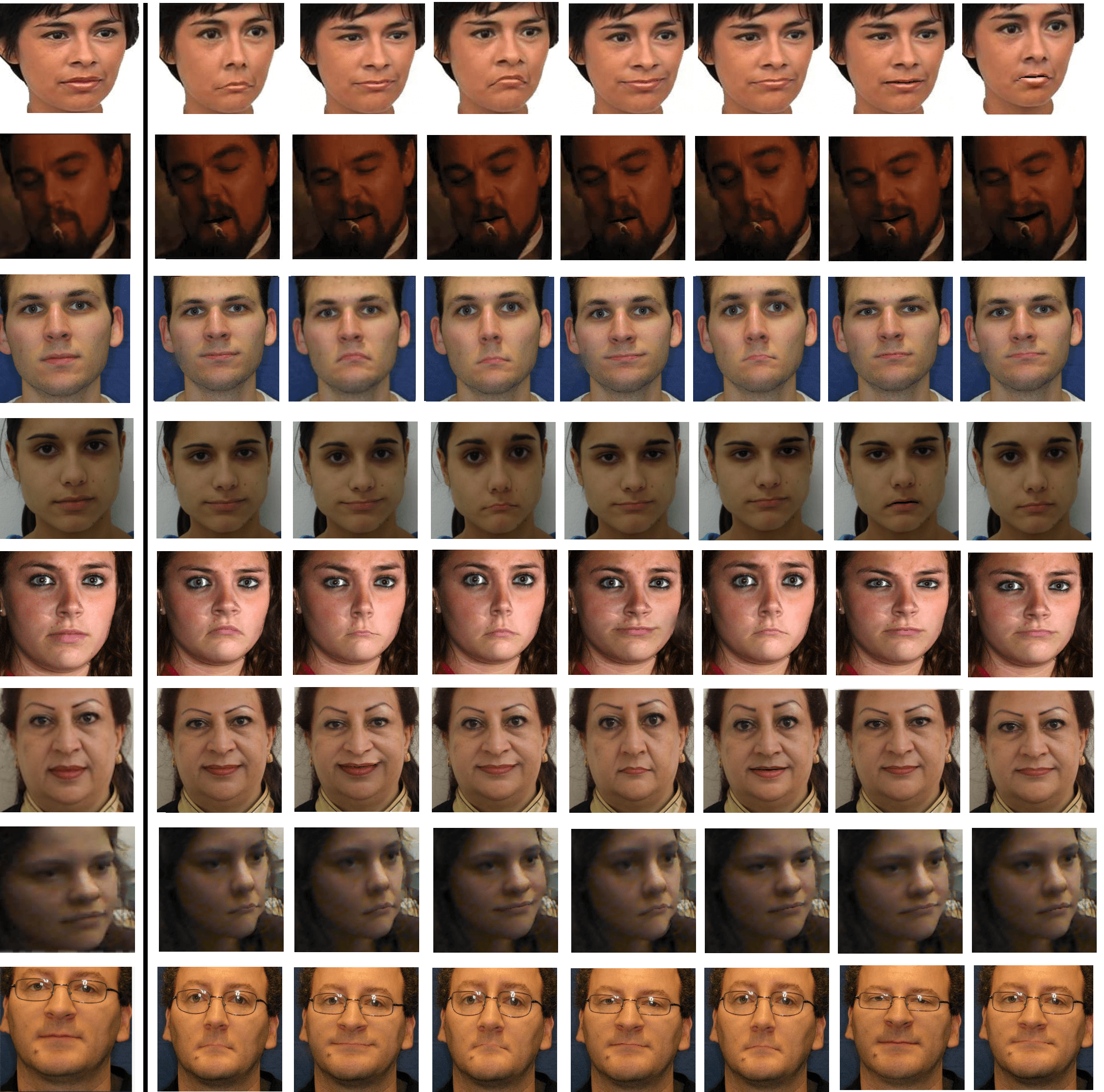} 
\caption{Synthesis of facial affect: on the left side are the neutral 2D images and on the right the synthesized images with different levels of affect}
\label{diff_expr}
\end{figure*}

Fig. \ref{results_all_dbs} is split into three parts. In each part, the top row illustrates some neutral images sampled from each of the aforementioned databases and the bottom one shows the respective synthesized images. Fig. \ref{diff_expr} shows the neutral images on the left side, and the synthesized images of different valence and arousal values on the right. It could be observed that our synthetic images are identity preserving, realistic and vivid. We showed that the proposed framework works well for images from both in-the-wild and controlled databases. This suggests that we could effectively synthesize facial affect regardless of different image conditions (e.g., occlusions, illumination and head poses).

\subsection{Quantitative evaluation of the facial affect synthesis}

\subsubsection{Leveraging synthetic data for training Deep Neural Networks} 
We used the synthetic faces to train deep neural networks for valence and arousal prediction on two facial affect databases annotated in terms of valence and arousal, the Aff-Wild and AFEW-VA. 
Our first step is to select neutral frames from these two databases.
Specifically, we selected frames with zero valence and arousal (human inspection was also conducted to make sure they are neutral faces), then, for each frame, we synthesized facial affect using the mean blendshape (as shown in Fig. \ref{blend_mean}) and assigned the median valence and arousal value of that class. 

\subsubsection{Experiments and data augmentation on the AFEW-VA} Following our approach, we created 108,864 synthetic images from the AFEW-VA database, a number that  is 3.5 times bigger than its original size. For training, we used the CNN-RNN (VGG-Face-GRU) architecture described in \cite{kollias2017recognition}.
Similarly to \cite{kossaifi2017afew}, we used a 5-fold person-independent cross-validation strategy and at each fold we augmented the training set with the synthesized images of people appearing only in that set (preserving the person independence). 
Table \ref{affewvacomp} shows a comparison of the performance of our network with the best results reported in \cite{kossaifi2017afew}. Those results are in terms of the Pearson Correlation Coefficient criterion (Pearson CC), defined as follows:

\begin{equation} \label{eq:1}
\rho_{xy} = \frac{s_{xy}}{s_x  s_y}
\end{equation}

\noindent
where $s_x$ and $s_y$ are the variances of the ground truth and predicted values respectively and $s_{xy}$ is the respective covariance value.

\begin{table}[h]
\caption{Pearson Correlation Coefficient evaluation of valence \& arousal predictions provided by the best architecture in \cite{kossaifi2017afew} vs the network trained on the augmented dataset created by our approach. Note that valence and arousal values are in $[-10,10$].}
\label{affewvacomp}
\centering
\begin{tabular}{ |c||c|c|c|c|  }
 \hline
 \multicolumn{1}{|c||}{Group} & \multicolumn{2}{|c|}{Pearson CC}  & \multicolumn{2}{|c|}{MSE} \\
 \hline
     & Valence & Arousal & Valence & Arousal  \\
\hline
best of \cite{kossaifi2017afew} & 0.407 & 0.45 & 6.96 & 4.97 \\
\hline
Our network (trained on the augmented dataset)  & \textbf{0.542} & \textbf{0.589}	& 4.75 &  2.74 \\
\hline
\end{tabular}
\end{table}

\subsubsection{Experiments and data augmentation on the Aff-Wild}: Following our approach, we created 60,135 synthetic images from the Aff-Wild database. We added those images to the training set of the first Affect-in-the-wild Challenge. It should be noticed that these images were synthesized from neutral faces found only in the training set of the challenge. The network we employed here was the the same CNN-RNN (VGG-Face-GRU) architecture described in \cite{kollias2017recognition}. 
Table \ref{table1} shows a comparison of the performance of our network trained with the augmented data with the best results reported in \cite{kollias2017recognition} and the results of the winner of the Aff-Wild Challenge \cite{weichi} (Method FATAUVA-Net).

\begin{table}[h]
\caption{Concordance Correlation Coefficient evaluation of valence \& arousal predictions provided by the CNN-RNN trained on the Aff-Wild dataset augmented with images synthesized by our approach vs methods \cite{weichi} \& \cite{kollias2017recognition}. Note that valence and arousal values are in $[-1,1$]. }
\label{table1}
\centering
\begin{tabular}{ |c||c|c|c|c| }
 \hline
 \multicolumn{1}{|c||}{} & \multicolumn{2}{|c|}{CCC} & \multicolumn{2}{|c|}{MSE}   \\
 \hline
     & Valence & Arousal & Valence & Arousal  \\
 \hline
FATAUVA-Net \cite{weichi}  & 0.396 & 0.282 & 0.123 & 0.095 		   \\
 \hline
\cite{kollias2017recognition} &0.570	&0.430 & 0.080 & 0.060	\\
 \hline
Our network trained on the augmented dataset & \textbf{0.591}&	\textbf{0.442} & 0.074 & 0.051	\\
 \hline 
\end{tabular}
\end{table}

From both tables, it can be verified that the network trained on the augmented, with synthetic images, dataset, outperformed the networks trained without them. This implies that, by augmenting the original training set, our methodology improved the network performance. It should be noted that the boost in performance is greater when the number of augmented images is much greater  than the number of images in the dataset (which is the case of AFEW-VA that contains 30,000 frames, while the augmented set included 109,000 more frames).

\section{Conclusions and Future Work}

A novel approach to generate facial affect in faces has been presented in this paper. It leverages a dimensional emotion model in terms of valence and arousal, and a large scale 4D face database, the 4DFAB. An efficient method has been developed for matching different blendshape models on large amounts of images extracted from the database and using these to render the appropriate facial affect on a selected face. A variety of faces and facial expressions has been examined in the experimental study, from ten databases showing expressions according to dimensional, but also categorical emotion models. The proposed approach has been successfully applied to faces from all databases, being able to render photorealistic facial expressions on them. 

In our future work we will extend this approach to synthesize, not only dimensional affect in faces, but also Facial Action Units. In this way a Global Local synthesis of facial affect will be possible, through a unified modeling of global dimensional emotion and local action unit based facial expression synthesis.

\subsubsection*{Acknowledgements.} The authors would also like to thank Evangelos Ververas for assisting with the affect synthesis.
The work of Dimitris Kollias was funded by a Teaching Fellowship of Imperial College London. 
The work of S. Cheng is funded by the EPSRC project EP/J017787/1 (4D-FAB) and EP/N007743/1 (FACER2VM).

\bibliographystyle{splncs04}
\bibliography{egbib}

\end{document}